\documentclass[aps,prl,amsmath,twocolumn]{revtex4}
 \renewcommand{\section}[1]{\paragraph{\textbf{#1}}}
\renewcommand{\section}[1]{\noindent\paragraph{\bf\emph{#1}}}

\newcommand{\AER}[1]{{\color{black} #1}}

	\usepackage{amsmath}
	\usepackage{amsfonts}
  	\usepackage{amssymb}
	\usepackage{makeidx}
	\usepackage{amsfonts}
	\usepackage[ansinew]{}
	\usepackage[usenames,dvipsnames]{pstricks}
	\usepackage{epsfig}

\usepackage{xcolor}

\newcommand{\be}{\begin{equation}}
\newcommand{\ee}{\end{equation}}
\newcommand{\bea}{\begin{eqnarray}}
\newcommand{\eea}{\end{eqnarray}}

\newcommand\SPD{\mathrel{\stackrel{\makebox[0pt]{\mbox{\normalfont\tiny (3)}}}{\Delta}}}

\newcommand\R{\zeta}	
\makeindex

\begin{document}

\title{The MESS of cosmological perturbations}
\author{Antonio Enea Romano$^{1,2}$, Sergio A. Vallejo Pe\~na$^{2}$}

\affiliation{
${}^{1}$Theoretical Physics Department, CERN, CH-1211 Geneva 23, Switzerland\\
${}^{2}$ICRANet, Piazza della Repubblica 10, I--65122 Pescara, Italy \\
}

\begin{abstract}
We introduce two new effective quantities for the study of comoving curvature perturbations $\R$: the \textit{space} dependent effective sound speed (SESS) and the \textit{momentum} dependent effective sound speed (MESS) . We use the SESS and the MESS to derive a new set of equations  which can be applied to  any system described by an effective stress-energy-momentum tensor (EST), including multi-fields systems, supergravity  and modified gravity theories. We show that this approach is completely equivalent to the standard one and it has the advantage of requiring to solve  \textit{only one} differential equation for $\R$ instead of a system, without the need of explicitly computing the evolution of entropy perturbations. 
The equations are valid for perturbations respect to any arbitrary flat spatially homogeneous background, including any inflationary and bounce model.

As an application we derive the equation for $\zeta$ for multi-fields  $KGB$ models and show that observed features of the primordial curvature perturbation  spectrum are compatible with the effects of an appropriate local variation of the MESS in momentum space. The MESS  is the \textit{natural} quantity to parametrize in a model independent way the effects produced on curvature perturbations by multi-fields systems, particle production and modified gravity theories and could be conveniently used in the analysis of LSS observations, such as the ones from the upcoming EUCLID mission  or  CMB radiation measurements. 
\end{abstract}


\maketitle
\section{Introduction}
The evolution of  comoving curvature perturbations $\R$ has a fundamental importance in cosmology since it is the basis for the study of different phenomena such as the CMB anisotropy or LSS. 
When the Universe is dominated by more than a scalar field the standard approach consists in deriving an equation for $\R$, with a time dependent sound speed, 
obtaining a new source term \cite{Gordon:2001} compared to the single field case, related to the field perturbations, which is interpreted as entropy perturbation.
In this paper we show that there exist a completely equivalent form of the equations which involves a space dependent effective sound speed (SESS) and no  source term. The advantage of this new approach is that it allows to study the evolution of curvature perturbations using only one equation for $\R$  without the need to introduce the notion of entropy perturbations or to integrate the system of differential equations for all the fields, as done in the standard approach \cite{Gordon:2001}. The same SESS can be defined for any system which in the standard formalism has entropy perturbations, including modified gravity theories where intrinsic entropy can be present also in the single field case, as in Horndeski's \cite{Horndeski:1974wa,prep} theory for example. We  show with some  examples that the new equation in terms of the SESS reduces to the equation with entropy terms, and derive the general relation between entropy and SESS. 

After defining the momentum dependent effective sound speed (MESS), an equation for $\R$ in momentum space is  derived and it is shown  that features of the spectrum of curvature perturbations can be explained as  variations of the MESS in momentum space.
We also derive the equation for  $\R$ for some single and multi-fields modified gravity scalar theories belonging to the Horndeski's family.

\section{Derivation of the equation for comoving curvature perturbations}
In order to derive an equation for comoving curvature perturbations $\R$ we  manipulate the perturbed Einstein's equations, assuming this general form for the scalar perturbations of the metric and of the EST
\begin{align}
ds^2 &= -(1+2A)dt^2+2a\partial_iB dx^idt+ \nonumber \\ & \quad {} + a^2\left\{\delta_{ij}(1+2C)+2\partial_i\partial_jE\right\}dx^idx^j \, , \label{pmetric} \\
T^0{}_0 &= - (\rho+\delta\rho) \quad \,, \quad  T^0{}_i = (\rho+P) \partial_i(v+B) \,, \nonumber \\ T^i{}_j &= (P+\delta P)\delta^i{}_j + \delta^{ik}
\partial_{k}\partial_{j}\Pi
-\frac{1}{3} \delta^{i}{}_{j} \SPD \Pi  \, . \label{psem}
\end{align}
where $v$ is the velocity potential and $\SPD \equiv \delta^{kl}\partial_k\partial_l$.

Note that the above equations are general, since no gauge has been chosen, and they can describe also multi-fields systems or modified gravity theories.
The comoving slices gauge, which we will call for brevity comoving gauge, is defined by the condition $(T^{0}{}_{i})_c=0$, where from now on we will be denoting with a subscript $c$ quantities evaluated on comoving slices. We will use this notation for the metric and the perturbed EST in the comoving gauge
\begin{align}
ds^2 &= -(1+2\gamma)dt^2+2a\partial_i\mu\, dx^idt+ \nonumber \\ & \quad {} + a^2\left\{\delta_{ij} (1+2\R)+2\partial_i\partial_j\nu\right\}dx^idx^j \, , \label{pmetricC} \\
(T^0{}_0)_c &= - (\rho+\beta) \, , \\ (T^i{}_j)_c&=(P+\alpha)\delta^i{}_j + \delta^{ik}
\partial_{k}\partial_{j}\Pi
-\frac{1}{3} \delta^{i}{}_{j} \SPD \Pi \,.
\end{align}
where we have defined the gauge invariant quantities $\alpha=\delta P_c,\beta=\delta \rho_c,\gamma=A_c,\mu=B_c,\zeta=C_c,\nu=E_c$.

In the case of a single scalar field with a canonical kinetic term or in K-inflation the \textit{comoving gauge} coincides with the \textit{uniform field gauge}, also known as \textit{unitary gauge}, but for multi-field systems they are \textit{different}. From the gauge transformation 
\begin{eqnarray}
v+B&\to&v+B- \delta t \, ,
\end{eqnarray}
we can derive the infinitesimal time translation taking to the comoving gauge, in which, $v_c+\mu=0$ 
\begin{eqnarray}
&\delta t_c=&  v+B \, .
\end{eqnarray}
We can then define the following gauge invariant quantities 
\begin{eqnarray}
\alpha&=& \delta P + \dot{P} \delta t_c \, \quad , \quad 
\beta = \delta \rho + \dot{\rho} \delta t_c \, , \\ 
\gamma &=& A + \dot{\delta t_c} \, \quad , \quad
\mu= B - a^{-1}\delta t_c \, , \\ 
\sigma&=& a \dot{E}-B+a^{-1}\delta t_c = a \dot{\nu}-\mu \, , \\ 
\R&=& C - H \delta t_c \, .
\end{eqnarray}
In the comoving gauge the Einstein's equations take the form \cite{Kodama:1985bj,Mukhanov:1990me} 
\begin{align}
 \frac{1}{a^2} \SPD [-\R+a H \sigma] &= \frac{\beta}{2}  \, , \label{eqzz}\\ 
\gamma &= \frac{\dot{\R}}{H}  \, , \label{eqzi} \\
-\ddot{\R}-3H\dot{\R}+H \dot{\gamma} +(2\dot{H}+3H^2)\gamma  &= \frac{ 1}{2} \left( \alpha + \frac{2}{3} \SPD \Pi \right) \, , \label{eqii} \\
\dot{\sigma}+2H\sigma-\frac{\gamma+\R}{a} &= a \Pi \, , \label{eqij}
\end{align}
where a dot denotes a partial derivative respect to cosmic time $t$ and we use a system of units in which $c=\hbar=M_{Pl}=1$.

After replacing eq.(\ref{eqzi}) into eq.(\ref{eqii}) we get the useful relation
\begin{align}
\dot{\R}= - \frac{1}{2H\epsilon} \left( \alpha +\frac{2}{3} \SPD \Pi \right) \, , \label{RdPPi}
\end{align}
where $\epsilon=-\dot{H}/H^2$. As we will see later this is the key equation to derive conservations laws for $\R$ and the second order differential equation in a closed form.
\AER{The quantities $\sigma$ and $\gamma$ are related to  the Bardeen potentials \cite{Bardeen:1980kt} $\Phi_B$ and $\Psi_B$ by}
\begin{align}
\sigma &= \frac{\Psi_B + \R}{aH} \quad , \quad \label{AB} 
\gamma = \Phi_B + \partial_t (a\sigma) \, . 
\end{align}
Replacing eq.(\ref{AB}) into eq.(\ref{eqzz}) we get  
\begin{align}
\frac{1}{a^2}\SPD \Psi_B &= \frac{1}{2} \beta \, . \label{Poisson}
\end{align}

We define the space dependent effective sound speed (SESS) of  comoving curvature perturbations according to the following relation
\begin{equation}
v_s^2 (t,x^i) \equiv \frac{\alpha (t,x^i)}{\beta (t,x^i)} \, , \label{vs}
\end{equation}
where $\alpha$ and $\beta$ are the perturbed pressure and energy density in the comoving gauge. 

We can now combine eqs.(\ref{RdPPi}), (\ref{Poisson}) and (\ref{vs}) to obtain
\begin{align}
 \dot{\R} &= -\frac{v_s^2}{a^2 H \epsilon} \SPD\Psi_B - \frac{1}{3H\epsilon } \SPD\Pi \, . \label{RPhiBPi} 
\end{align}
 Replacing eq.(\ref{AB}) into eq.(\ref{eqzi}) we get the useful relation between the comoving curvature perturbation and the Bardeen potentials 
\begin{align}
\R &= - \Psi_B + \frac{H^2}{\dot{H}} \left( \Phi_B + H^{-1}\dot{\Psi}_B\right)  \, .   \label{RB}
\end{align}
After substituting eq.(\ref{AB}) into eq.(\ref{eqij}) we get 
\begin{equation}
\Phi_B = \Psi_B - a^2 \Pi \, ,
\end{equation}
which we can replace into eq.(\ref{RB}) to obtain
\begin{equation}
\R = - \frac{1}{a \epsilon} \partial_t \left( \frac{a^3 \Psi_B}{H} \right) + \frac{a^2}{\epsilon} \Pi \, ,
\end{equation}
and combining it with the time derivative of eq.(\ref{RPhiBPi})  we finally get 
\begin{align}
\partial_t \left( Z^2 \dot{\R} \right) &-  a \epsilon \SPD \R + a^3 \SPD \Pi + \partial_t \left( \frac{Z^2}{3 H \epsilon} \SPD \Pi \right) =0 \, ,   \label{RceqPi} 
\end{align}
where $Z^2\equiv\epsilon a^3/v_s^2$. We can also write this equation as \begin{align}
\ddot{\R} + \frac{\partial_t(Z^2)}{Z^2} \dot{\R} &- \frac{v_s^2}{a^2} \SPD \R + \frac{v_s^2}{\epsilon}\SPD \Pi  + \nonumber \\ & \quad {}+  \frac{1}{3 Z^2}\partial_{t}\left( \frac{Z^2}{H \epsilon} \SPD \Pi \right)= 0 \, . \label{RcttPi}
\end{align}
Equations  (\ref{RPhiBPi}), (\ref{RceqPi}), and (\ref{RcttPi}) together with the momentum space form derived later in eq.(\ref{Rckeq}-\ref{ukeq}) are the main theoretical results of this paper. Note that in absence of anisotropy eq.(\ref{RPhiBPi}) is similar to the one obtained in \cite{Bellido:1995}, but without the entropy terms. We will explain this difference in more details in the section about two scalar fields, and more in general in the section about the relation between SESS and entropy.
 
\AER{Quite remarkably  the equations above can be applied to any system described by an appropriate EST, including multi-fields, supergravity, and  modified gravity theories.}

The equation is also valid for example for single field models with intrinsic entropy \cite{prep}, such as the $KGB$  theories \cite{Deffayet:2010qz}, or other  Horndeski's theories.

\section{Relation with entropy}
In the standard approach \cite{Kodama:1985bj} entropy perturbations $\Gamma$ are defined by
\begin{align}
\alpha(t,x^i) &= c_s(t)^2 \beta(t,x^i) + \Gamma(t,x^i) \, , \label{entropy} \end{align}
where $c_s$ is interpreted as sound speed and is a function of time only.
This definition is invariant under the transformation
\begin{align}
c_s^2 & \to \tilde{c}_s(t)^2=c_s(t)^2+\Delta c_s(t)^2 \, , \label{cst} \\ 
\Gamma & \to \tilde{\Gamma}(t,x^i) = \Gamma(t,x^i) - \Delta c_s(t)^2 \beta (t,x^i) \, , \label{Gammat}
\end{align}
where $\Delta c_s(t)^2$ is an arbitrary function of time only. The invariance of eq.(\ref{entropy}) under these transformations shows that this definition of entropy is not unique.
\AER{This \textit{ambiguity} in the definition of the sound speed $c_s(t)$ and of the entropy perturbations is another motivation to introduce the SESS, which on the contrary is a \textit{uniquely} defined gauge invariant quantity.} 

Combining eq.(\ref{entropy}), (\ref{vs}) and (\ref{RdPPi}) we get the relation between SESS and entropy 
\begin{align}
v_s^2 &= c_s^2\left[ 1 + \frac{\Gamma}{2 H \epsilon \left(\dot{\R} + \frac{1}{3H\epsilon} \SPD \Pi\right)}\right]^{-1} \, . \label{vcgamma}
\end{align}
Replacing eq.(\ref{vcgamma}) into eq.(\ref{RPhiBPi}) and eq.(\ref{RceqPi})  we get
\begin{align}
\dot{\R} &= - \frac{c_s^2}{a^2H\epsilon} \SPD\Phi_B - \frac{\Gamma}{2H\epsilon} - \frac{1}{3H \epsilon} \SPD \Pi\, , \label{RPhiGamma} \\ 
\ddot{\R}&+\frac{\partial_t z^2}{z^2}\dot{\R}-\frac{c_s^2}{a^2}\SPD\R + \frac{c_s^2}{\epsilon}\SPD \Pi + \nonumber \\ & \quad {} + \frac{1}{z^2}\partial_t \left[ \frac{a^3}{c_s^2 H} \left(\Gamma + \frac{2}{3}\SPD \Pi \right) \right] = 0 \, , \label{Rdotcgamma}
\end{align}

where $z^2=2a^3\epsilon/c_s^2$.
These equations are in agreement with the  equation derived \cite{Vallejo1:2018} using the standard definition of entropy perturbation given in eq.(\ref{entropy}).
\section{Application to two scalar fields}
In order to show that eq.(\ref{RPhiBPi}) and eq.(\ref{RceqPi}) are valid also for multi-fields systems we will consider the example of  two scalar fields minimally coupled to gravity with Lagrangian
\begin{align}
L = \sum^2_n -X_n-2V(\Phi_1,\Phi_2)  \, , \nonumber
\end{align}
where $X_n=g^{\mu\nu}\partial_\mu\Phi_n\partial_\nu\Phi_n$.
In this section we will use this notation $\Phi_n(x^{\mu})=\phi_n(t) + \delta \phi_n(x^{\mu}),\Phi=\Phi_1,\Psi=\Phi_2$.
The components of the perturbed EST of the two scalar fields system, without gauge fixing, are 
\begin{align}
\delta T^{0}{}_{0} &= -\dot{\phi} \dot{\delta\phi} - \dot{\psi} \dot{\delta\psi} + A (\dot{\phi}^2+\dot{\psi}^2) - V_{\phi} \delta \phi - V_{\psi} \delta \psi \nonumber \, ,  \\
\delta T^{i}{}_{j} &= \delta^{i}_{j} \left[ \dot{\phi} \dot{\delta\phi} + \dot{\psi} \dot{\delta\psi} - A (\dot{\phi}^2+\dot{\psi}^2) - V_{\phi} \delta \phi - V_{\psi} \delta \psi \right] \,  \nonumber  \\
\delta T^{0}{}_{i} &= \partial_i \left( -\frac{\dot{\phi} \delta \phi+\dot{\psi} \delta \psi}{a} \right) \, , \label{pq}
\end{align}
where we are denoting the partial derivatives as $V_{\phi}=\partial_{\phi} V(\phi,\psi)$ and $V_{\psi}=\partial_{\psi} V(\phi,\psi)$.  

The comoving gauge is defined by the condition 
$(\delta T^{0}{}_{i})_c=0$, which implies that 
\begin{equation}
\dot{\phi}U_{\phi}+\dot{\psi}U_{\psi} = 0 \, ,
\end{equation}
where  $U_n=(\delta\phi_n)_c=\delta\phi_n+\dot{\phi}_n\delta t_c$ are the field perturbations in the comoving gauge, as defined in eq.(\ref{cgt}), which are by construction gauge invariant. It is important to note that, contrary to the single field case, the \textit{comoving gauge} for multi-fields systems \textit{does not coincide} with the \textit{uniform field gauge}, also known as \textit{unitary gauge}, since there is not enough gauge freedom to set both  fields perturbation to zero, i.e. in general it is impossible to use gauge transformation to choose a coordinate system in which $\delta\phi=\delta\psi=0$. This is the origin of the  \textit{space dependency} of the SESS for multi-fields systems.

Under an infinitesimal time translation $t \to t-\delta t$ the fields perturbations transform according to the gauge transformation
\begin{align}
\widetilde{\delta \phi} &= \delta \phi + \dot{\phi} \delta t \quad , \quad \widetilde{\delta \psi} = \delta \psi + \dot{\psi} \delta t \, . \label{gtdeltapsi}
\end{align}
From these equations we can find the time translation $\delta t_c$ necessary to go to the comoving gauge, by imposing the comoving gauge condition $(\delta T^{0}{}_{i})_c=0 \rightarrow \dot{\phi}\delta\widetilde{\phi}+\dot{\psi}\delta\widetilde{\psi} = 0$,  obtaining 
\begin{align}
\delta t_c &= -\frac{\dot{\phi} \delta \phi+\dot{\psi} \delta \psi}{\dot{\phi}^2+\dot{\psi}^2} \, . \label{cgt}
\end{align}

We can now compute explicitly gauge invariant quantities defined in the comoving gauge, such as the comoving field perturbation 
\begin{align}
U_{\phi} &= \delta \phi - \dot{\phi} \frac{\dot{\phi} \delta \phi+\dot{\psi} \delta \psi}{\dot{\phi}^2+\dot{\psi}^2} \, , \, U_{\psi} = \delta \psi - \dot{\psi} \frac{\dot{\phi} \delta \phi+\dot{\psi} \delta \psi}{\dot{\phi}^2+\dot{\psi}^2} \, , \nonumber
\end{align}
and the comoving  pressure and energy perturbations 
\begin{align}
\alpha &=\delta P_c= \dot{\phi} \dot{U}_{\phi} + \dot{\psi} \dot{U}_{\psi} - \gamma (\dot{\phi}^2+\dot{\psi}^2) + \nonumber \\ &  \quad{} + (\ddot{\phi} + 3 H \dot{\phi}) U_{\phi} +(\ddot{\psi} + 3 H \dot{\psi}) U_{\psi} \, ,\\
\beta &=\delta \rho_c= \dot{\phi} \dot{U}_{\phi} + \dot{\psi} \dot{U}_{\psi} - \gamma (\dot{\phi}^2+\dot{\psi}^2) + \nonumber \\ &  \quad{} - (\ddot{\phi} + 3 H \dot{\phi}) U_{\phi} -(\ddot{\psi} + 3 H \dot{\psi}) U_{\psi} \label{deltarhocT} \,.
\end{align}
After replacing eq.(\ref{eqzi}) and the expressions for $U_n$ into these equations we finally get  
\begin{align}
\beta &= - \frac{\dot{\R}(\dot{\phi}^2+\dot{\psi}^2)}{H} -\frac{\Theta (\dot{\phi}^2+\dot{\psi}^2)}{2} \, , \, \alpha = - \frac{\dot{\R}(\dot{\phi}^2+\dot{\psi}^2)}{H} \, ,  \nonumber
\end{align}
where the function $\Theta$ is defined according to
\begin{align}
\Theta &= \left(\frac{\delta\phi}{\dot{\phi}} - \frac{\delta\psi}{\dot{\psi}} \right) \frac{\partial}{\partial t} \left( \frac{\dot{\phi}^2-\dot{\psi}^2}{\dot{\phi}^2+\dot{\psi}^2} \right) \, . \label{tft}
\end{align}
It is important to note that the above expression is gauge invariant,  i.e.
\begin{equation}
\left(\frac{\delta\phi}{\dot{\phi}} - \frac{\delta\psi}{\dot{\psi}} \right) = \left(\frac{Q_{\phi}}{\dot{\phi}} - \frac{Q_{\psi}}{\dot{\psi}} \right)= \left(\frac{U_{\phi}}{\dot{\phi}} - \frac{U_{\psi}}{\dot{\psi}} \right)
\end{equation}
where 
\begin{align}
Q_{\phi} & \equiv \delta \phi + \frac{\dot{\phi}}{H} C \, \quad, \quad
Q_{\psi}  \equiv \delta \psi + \frac{\dot{\psi}}{H} C \, , \label{Qpsi}
\end{align}
are the field perturbations in the flat gauge.

Assuming a classical field trajectory parametrized as $\psi(\phi)$ we can write $\Theta$  in this form
\begin{eqnarray}
\Theta=4\dot{\phi} \frac{\partial\psi}{\partial\phi}  \frac{\partial^2\psi}{\partial\phi^2}\left[\left(\frac{\partial\psi}{\partial\phi}\right)^2+1\right]^{-2}
\left(\frac{U_{\psi}}{\dot{\psi}} - \frac{U_{\phi}}{\dot{\phi}} \right) 
\end{eqnarray}
From the above expression we can see that in order for $\Theta$ to be different from zero the trajectory has to have non vanising first and second derivatives, i.e. there must be some turn in the field space.

Combining eqs.(\ref{vs}) with the expressions derived above for $\alpha$ and $\beta$ we  obtain  the SESS for the two scalar fields
\begin{align}
v_s^2 &= \left(1+\frac{H \Theta}{2 \dot{\R}}\right)^{-1} \, . \label{ss} 
\end{align}
and replacing it  into eq.(\ref{RPhiBPi}) and eq.(\ref{RcttPi}) we get 
\begin{align}
\dot{\R} &=  \frac{H}{a^2 \dot{H}} \SPD \Phi_B - \frac{1}{2} H \Theta \, , \label{RdotPhi}   \\ 
\ddot{\R}&+\frac{\partial_t (z^2)}{z^2}\dot{\R}-\frac{1}{a^2}\SPD\R+\frac{1}{z^2}\partial_t \left( \frac{z^2H\Theta}{2} \right)=0 \, , \label{Rceqs}
\end{align}
in agreement with \cite{Finelli:2000ya,Bellido:1995,Gordon:2001}, confirming that eq.(\ref{RPhiBPi}), obtained for a generic EST, is indeed valid also for multi-field systems. 
When the relation between the SESS and $\dot \R$ in eq.(\ref{ss}) is used explicitly the equation for $\R$ has a source term which is commonly interpreted as entropy \cite{Gordon:2001,Bellido:1995}, while in the form of eq.(\ref{RcttPi}) there is no entropy but the SESS is not only time but also space dependent. This example shows that the two descriptions are completely equivalent, and that the notion of entropy is in fact not really necessary, as long as the SESS is defined appropriately as in eq.(\ref{vs}), as a \textit{space-time} dependent quantity.

The evolution of comoving curvature perturbations for multiple scalar fields systems could  be studied without introducing any notion of entropy perturbation, by  specifying an appropriate SESS $v_s(x^{\mu})$.
The advantage of this approach is that SESS can be used as an effective quantity without explicitly specifying the multi-fields model or solving the system of differential equations for the field perturbations $Q_{\phi}$ and $Q_{\psi}$, as done in  the standard approach \cite{Gordon:2001}.

As we have shown for the particular case of two fields, for an arbitrary number of scalar fields  it is enough to solve just equation (\ref{RcttPi}), and this can be a substantial computational advantage in presence of several fields.
\section{The SESS for generic multi-fields systems and particle production}
The Lagrangian for N fields system with standard kinetic term is
\begin{align}
L &=  \sum_n^N -X_n - 2 V(\Phi_n) \,  \label{actiont}
\end{align}
where $X_n=g^{\mu\nu}\partial_\mu\Phi_n\partial_\nu\Phi_n$ and $\Phi_n(x^{\mu}) = \phi_n(t) + \delta \phi_n(x^{\mu})$.
Depending on the type of  potential, particle production could occur \cite{Romano:2008rr,Chung:1999ve} and this would correspond to a specific form of the SESS $v_s(x^{\mu})$.  The effects of particle production on curvature perturbations \cite{Romano:2008rr,Chung:1999ve} could be modeled phenomenologically, without specifying the potential or the number of field, by considering different forms of the SESS $v_s(x^{\mu})$.
The results derived in the previous section can be generalized to
\begin{align}
v_s^2 &= \left(1+\frac{H \Theta}{2 \dot{\R}}\right)^{-1} \, . \label{sst} 
\end{align}
where the function $\Theta$ is  proportional to
\begin{align}
\theta_{ij}=&\left(\frac{\delta\phi_i}{\dot{\phi_i}} - \frac{\delta\phi_j}{\dot{\phi_j}} \right) \frac{\partial}{\partial t} \left( \frac{\dot{\phi_i}^2-\dot{\phi_j}^2}{\sum_{i}^{n} \dot{\phi_i}^2} \right) \,,\, \Theta = \chi_N \sum_{i>j}^N \theta_{ij} \nonumber
\end{align}
and  $\chi_N$ is an appropriate  combinational factor depending on the number of fields $N$.

Comparing with eq.(\ref{vcgamma}) we get that $\Gamma=\epsilon H\Theta$, from which we can define $\Gamma_{ij}=\epsilon H\Theta_{ij}$ as the entropy due to the interaction between fields pairs, in terms of which the total interacting entropy is given by $\Gamma = \chi_N \sum_{i>j}^n \Gamma_{ij}$.
\section{Application to modified gravity theories}
\AER{For modified gravity theories the EST is defined by writing the field equations as $G_{\mu\nu}=T^{eff}{\mu\nu}$}, where $G_{\mu\nu}$ is the Einstein's tensor. For single field Horndeski theory the decomposition of the perturbed EST shows the presence of intrinsic entropy $\Gamma^{int}$ \cite{prep}
\begin{align}
\alpha=c_s^2(t)\beta+\Gamma^{int}
\end{align}
and in the particular case of KGB models with Lagrangian \cite{Deffayet:2010qz}
\begin{align}
L_{KG}(\Phi,X) &= K\left(\Phi,X\right) + G\left(\Phi,X\right) \Box \Phi  \, , \nonumber  
\end{align}
there is no anisotropy, so that equation (\ref{vcgamma}) is valid with this definition of SESS
\begin{eqnarray}
v^2_{KG}=c_{s}^2\left( 1 + \frac{\Gamma^{int}}{2\epsilon H \dot{\R}}\right)^{-1} \, .
\end{eqnarray}

The same result could also be generalized to multi-field KGB models with Lagrangian  $L_{NKG}=\sum_n^N L_{KG}(\Phi_n,X_n)$, in which case the total entropy would be given by the sum of the intrinsic and interaction entropy as
\begin{eqnarray}
\Gamma_{NKG}&=&\sum_i^N \Gamma_i^{int}+\chi_N \sum_{i>j}^N\Gamma_{ij} \, , 
\end{eqnarray}
and the corresponding SESS is
\begin{eqnarray}
v^2_{NKG}&=&c_{s}^2\left( 1 + \frac{\Gamma_{NKG}}{2\epsilon H \dot{\R}}\right)^{-1} \,.
\end{eqnarray}
\section{Momentum dependent effective sound speed}
We define the momentum dependent effective sound speed (MESS) $\tilde{v}_k(t)^2$ as
\begin{equation}
\tilde{v}_k^2(t) \equiv \frac{\alpha_k(t)}{\beta_k(t)} \, , \label{ck}
\end{equation}
and following a procedure mathematically similar to the one used to derive eq.(\ref{RcttPi}) we can obtain this equation in momentum space \cite{prep}
\begin{align}
\ddot{\R}_k + \frac{\partial_t(\tilde{Z}_k^2)}{\tilde{Z}_k^2} \dot{\R}_k & + \frac{\tilde{v}_k^2}{a^2} k^2 \R_k - \frac{\tilde{v}_k^2}{\epsilon}k^2 \Pi_k + \nonumber \\ & \quad {}-  \frac{1}{3 \tilde{Z}_k^2}\partial_{t}\left( \frac{\tilde{Z}_k^2}{H \epsilon} k^2 \Pi_k \right)= 0  \, , \label{Rckeq}  
\end{align}
where $ \tilde{Z}_k^2\equiv\epsilon a^3/\tilde{v}_k^2$.
It is important to note that the MESS $\tilde{v}_k(t)$  defined in eq.(\ref{ck}) is not simply the Fourier transform of the SESS $v_s(x^{\mu})$ defined in eq.(\ref{vs}), because the product of the  Fourier transforms of two functions is the transform of the convolution of the two functions.
\section{The effects of local variation of the MESS}
As an example let's consider the case of an isotropic EST for which, after introducing the variable $u_k\equiv \tilde{Z}_k \R_k$, eq.(\ref{Rckeq}) takes the form
\begin{equation}
\ddot{u}_k + \left(\frac{\tilde{v}_k^2 k^2}{a^2} - \frac{\ddot{\tilde{Z}}_k}{\tilde{Z}_k} \right) u_k =0 \, , \label{ukeq}
\end{equation}
which reduces to the Sasaki-Mukhanov equation when $\tilde{v}_k$ is a function of \textit{time only}.
In the case in which $\tilde{v}_k$ is \textit{not} a function of time, equations (\ref{Rckeq}) and (\ref{ukeq}) can be written as
\begin{align}
\R_k''+\frac{\partial_{\eta}(z^2)}{z^2}\R_k'+\tilde{v}_k^2k^2\R_k &=0 \, , \label{Rcpk} \\
u_k''+\left(\tilde{v}_k^2 k^2 -\frac{z''}{z} \right)u_k&=0 \, , \label{ukp}
\end{align}
where the prime denotes derivatives with respect to conformal time $\tau$, and $z^2=2a^2\epsilon$.

\begin{figure}
\includegraphics[width=0.5\textwidth]{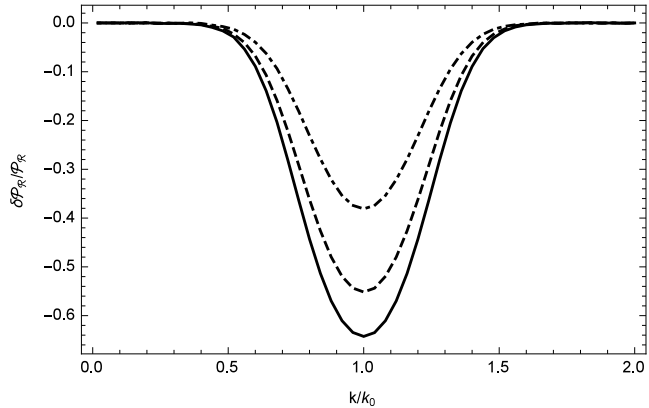}
\caption{The relative difference $\Delta \mathcal{P}_{\R}/\mathcal{P}_{\R}$ is plotted as a function of  $k/k_0$. The solid, dashed and dot-dashed lines correspond $\sigma=2.5 \times 10^{-1} k_0$ and  $A_c=4\times 10^{-1}$, $A_c=3\times 10^{-1}$ and $A_c=1.7\times 10^{-1}$ respectively.}
\label{deltaPR}
\end{figure}

One interesting application of this equation consists in computing the effects of the MESS on the spectrum of primordial curvature perturbations, motivated by the observations pointing to local deviations from power law \cite{Benetti:2012wu,Horiguchi:2017ume,Hunt:2015iua,Ade:2015lrj}. 
In fig.(\ref{deltaPR}) we plot the relative difference $\Delta P_{\R}/P_{\R}$ between the power spectrum of a slow-roll model compatible with observations and with $\tilde{v}_k=1$, and that of a model with the same parameters, and the MESS given by
\begin{align}
\tilde{v}_k = 1 + A_c \exp \left[ -\left( \frac{k-k_0}{\sigma} \right)^2 \right] \, , \label{ckg}
\end{align}
where  $k_0$ is an arbitrary scale, for example it could be $k_0\approx 1.5 \times 10 ^{-3} Mpc^{-1}$, which corresponds to the location of one of the most statistically significant features  \cite{Ade:2015lrj}. 

\AER{The ansatz in eq.(\ref{ckg}) is phenomenological, motivated by the theoretical interest in computing the effects of a local variation of the MESS on the curvature perturbations spectrum. Different types of physical models could produce it, such as multi-fields or modified gravity theories, but here we focus more on its phenomenological effects rather than on its fundamental origin, leaving this investigation to future works. In particular  we are interested in clarifying how the MESS can produce the super-horizon evolution of curvature perturbations which in the standard approach is attributed to the entropy source term. 
In this regard we can note that due to the non trivial MESS the modes which leave the horizon around $\tau_0=-1/k_0$ do not freeze right after but can have some residual super-horizon evolution, because the freezing time is $\tau_f=-1/\tilde{v}_k k$, which is different from the horizon crossing time $\tau_H=-1/k$.  Because of this difference  the curvature perturbations modes can have a super-horizon evolution despite of the absence of a source term in eq.(\ref{ukp})}.

We compute the spectrum choosing the Bunch-Davies vacuum \cite{Bunch:1978yq} for modes deeply inside the horizon, corresponding to
\begin{align}
\R_k &= \frac{1}{2a\sqrt{\epsilon \tilde{v}_k k}} e^{-i\tilde{v}_k k\tau} \, , \label{MLMSR}
\end{align}
which has been normalized using the Wronskian condition for $u_k$. 

As shown in fig.(\ref{deltaPR}) the effects of the MESS on the curvature perturbation spectrum are qualitatively in agreement with the local features whose presence is supported by observations. A more systematic analysis could consist in fitting observational data  using a  general parametrization for the MESS, for example a piecewise cubic Hermite interpolating polynomial(PCHIP). Note that in general the MESS can be both momentum and time dependent, and from a phenomenological point of view another convenient parameter to use for a model independent analysis is $\tilde{Z}_k(\tau)$, which could be fitted with a double PCHIP, one for the time and the other for the momentum dependence, corresponding for example to an ansatz of the form $\tilde{Z}_k(\tau)=f(k)g(\tau)$.

The advantage of this approach is that it allows to study the effects of a wide class of theoretical models such as multi-fields systems, modified gravity theories or a combination of the two such as for the $nKGB$ model discussed previously, using just one equation. Once the MESS has been constrained by model independent analysis of observational data the theoretical models can be compared to it.

\section{Conclusions}
After introducing a new definition of the space dependent effective sound speed (SESS) and the momentum dependent effective sound speed (MESS) we have derived a set of new equations (\ref{RPhiBPi}-\ref{RcttPi}) and (\ref{ukp}-\ref{ukeq}) for the comoving curvature curvature perturbations valid for any system described by an EST, including multi-fields systems, modified gravity theories, or a combination of the two. 
The approach we have developed is completely equivalent to the standard one, and it \textit{does not require any notion of entropy perturbation}.
\AER{
The main advantages for using the SESS and the MESS are
\begin{itemize}
\item they simplify significantly the phenomenological study of the evolution of curvature perturbations, reducing it to the solution of only one differential equation instead of a system
\item they are \textit{unique} gauge invariant quantities contrary to the standard definition of sound speed and entropy in eq.(\ref{entropy}), which does not completely fix them
\item they allow to make a model independent analysis of observational data using effective quantities  without specifying the details of the underlying theoretical model.
\end{itemize}
}
As an application we have shown that  features of the primordial curvature perturbation spectrum can be modeled with an appropriate choice of the MESS, and we have derived the equation for comoving curvature perturbations for a single field and multi-field modified gravity theory, the $nKGB$ model.
In the future it will be interesting to apply this new approach to the analysis of LSS observations and CMB data using a  PCHIP parametrization for both the time and space dependency of the MESS.

\section{Acknowledgments}
We thank Juan Garcia Bellido for helpful discussions. We also thank the anonymous referee for comments and suggestions.

\bibliographystyle{h-physrev4}
\bibliography{mybib}
\end{document}